\documentclass[8.5pt,twoside,twocolumn]{article}
\usepackage[super,sort&compress,comma]{natbib}
\usepackage[version=3]{mhchem}
\usepackage{balance}
\usepackage{times,mathptm}
\usepackage{sectsty}
\usepackage{graphicx} 
\usepackage{lastpage}
\usepackage[format=plain,justification=raggedright,singlelinecheck=false,font=small,labelfont=bf,labelsep=space]{caption} 
\usepackage{fancyhdr}
\begin{document}

\thispagestyle{plain}
\fancypagestyle{plain}{
\renewcommand{\headrulewidth}{1pt}}
\renewcommand{\thefootnote}{\fnsymbol{footnote}}
\renewcommand\footnoterule{\vspace*{1pt}%
\hrule width 3.4in height 0.4pt \vspace*{5pt}} 
\setcounter{secnumdepth}{5}

\makeatletter 
\renewcommand\@biblabel[1]{#1}            
\renewcommand\@makefntext[1]%
{\noindent\makebox[0pt][r]{\@thefnmark\,}#1}
\makeatother 
\renewcommand{\figurename}{\small{Fig.}~}
\sectionfont{\large}
\subsectionfont{\normalsize} 

\fancyfoot{}
\fancyfoot[RO]{\footnotesize{\sffamily{1--\pageref{LastPage} ~\textbar  \hspace{2pt}\thepage}}}
\fancyfoot[LE]{\footnotesize{\sffamily{\thepage~\textbar\hspace{3.45cm} 1--\pageref{LastPage}}}}
\fancyhead{}
\renewcommand{\headrulewidth}{1pt} 
\renewcommand{\footrulewidth}{1pt}
\setlength{\arrayrulewidth}{1pt}
\setlength{\columnsep}{6.5mm}
\setlength\bibsep{1pt}

\twocolumn[
  \begin{@twocolumnfalse}
\noindent\LARGE{\textbf{Oscillatory settling in wormlike-micelle solutions:
bursts and a long time scale}}
\vspace{0.6cm}

\noindent\large{\textbf{Nitin Kumar,\textit{$^{a}$} Sayantan Majumdar,\textit{$^{a}$} Aditya Sood,\textit{$^{b}$} Rama Govindarajan,\textit{$^{c}$} Sriram Ramaswamy \textit{$^{a}$} and
A.K. Sood\textit{$^{a}$}}}\vspace{0.5cm}

\noindent\textit{\small{\textbf{Received 11th January 2012, Accepted 14th February 2012\newline
Published in Soft Matter}}}

\noindent \textbf{\small{DOI: 10.1039/c2sm25077b}}
 \end{@twocolumnfalse} \vspace{0.6cm}

  ]

\noindent\textbf{We study the dynamics of a spherical steel ball falling freely through a
solution of entangled wormlike micelles. If the sphere diameter is larger than a
threshold value, the settling velocity shows repeated short oscillatory bursts
separated by long periods of relative quiescence. We propose a model
incorporating the interplay of settling-induced flow, viscoelastic stress and,
as in [PRE \textbf{66}, 025202(R) (2002) and PRE \textbf{73}, 041508 (2006)], a
slow structural variable for which our experiments offer independent
evidence.}
\section*{}
\vspace{-1cm}


\footnotetext{\textit{$^{a}$~Department of Physics, Indian Institute of Science, Bangalore 560 012, India}}
\footnotetext{\textit{$^{b}$~Department of Materials Science and Engineering, Stanford University
Stanford, CA 94305}}
\footnotetext{\textit{$^{c}$~Engineering Mechanics Unit, Jawaharlal Nehru Centre for Advanced Scientific Research, Bangalore 560 064, India. }}



A sphere falling through a Newtonian fluid at low Reynolds number Re
\cite{stokes} reaches a steady terminal velocity through the interplay of
gravity, buoyancy and viscous drag. This simple system, and its extensions to
unsteady high Re flow, are well understood and form the basis of falling ball
viscometers \cite{Chhabra}. In non-Newtonian fluids unsteady settling is
observed \cite{Mollinger, Belmonte} even at low Re. Mollinger \textit{et al.}
\cite{Mollinger} observed periodic stick-slip settling in the beginning of
the fall and investigated the effect of wall confinement on the baseline
settling velocity and the frequency of unsteady events. Jayaraman \textit{et
al.} \cite{Belmonte} found a transition from steady to \textit{irregular
unsteady} settling, remarked that the relevant control parameter was the shear
rate, $\dot{\gamma} = v_{b}/d$ ($v_{b}$ \& $d$ being the baseline velocity and
sphere diameter respectively), and emphasized the importance of the negative
wake behind the particle \cite{Harlen,Hassager-Bisgard}. It was suggested
\cite{Belmonte} that the oscillations were a result of the formation and breakup
of flow-induced structures, known to arise in concentrated micellar systems
\cite{Walker,Rehage,Cates_Fielding_review}.

\begin{figure}[h]
\centering
	\includegraphics[height=8.5cm]{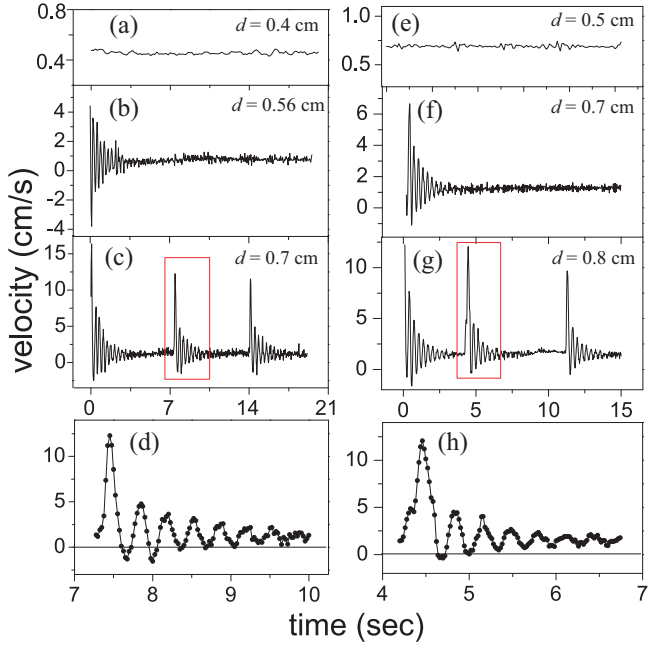}
	\caption{Velocity as a function of time: For Gel A; (a) d=0.4 cm, (b) d=0.56 cm and (c) d=0.7 cm. In (d), the velocity variation marked in (c) has been expanded. For Gel B: (e) d=0.5 cm, (f) d=0.7 cm and (g) d=0.8 cm. In (h), the velocity variation marked in (g) has been expanded.}\label{bursts}
\end{figure}

In this paper we investigate the sedimentation of spherical metallic particles
through an entangled wormlike-micelle solution \cite{Cates_Fielding_review} of
the surfactant CTAT (Cetyl Trimethyl Ammomium p-Toluenesulphonate) in brine. We
report here for the first time that spheres larger than a critical size not only
undergo unsteady motion, but also show sustained, repeated bursts of
oscillations superposed on a constant baseline velocity as shown in Fig.
\ref{bursts}. The intra-burst oscillation period $\tau_{fast} \sim d/v_b$; the
time between bursts $\tau_{slow}$ is at least an order of magnitude larger. We
account for these results through a model incorporating shear-induced coupled
oscillations of viscoelastic stress components and a slow structural relaxation
\cite{ajdarietal2002,aradiancates} giving rise to $\tau_{slow}$. We suggest that
equilibrium fluctuations, i.e., without imposed flow, should show signatures of
the structural variable.

Experiments were conducted on two different samples, Gel A comprising 2 wt$\%$ CTAT  $+$ 100 mM NaCl and Gel B with 2.2 wt$\%$ CTAT $+$ 82 mM NaCl. Rheological measurements were carried out on a Paar Physica MCR 300 Rheometer in the cone-plate geometry. The frequency sweep and flow curve of Gel A give an extrapolated zero-shear-rate viscosity of $17.8$ Pa s at 25$^{\circ}$C, onset of shear thinning at a shear rate $\dot{\gamma_c} \sim 0.4 s^{-1}$. The linear viscoelasticity is Maxwellian with a relaxation time $\omega_{co}^{-1} \simeq 1.34$ s. The corresponding numbers for Gel B are $8.5$ Pa s, 1.5 s$^{-1}$ and $0.97$ s respectively. These systems are in the parameter range where shear banding instabilities and spatiotemporal rheochaos are expected above a critical shear rate \cite{Ranjini, Rajesh, RajeshPRE,RajeshJNNFM}.

A transparent cylindrical tube of diameter $D=$ 5.4 cm and height 40 cm was filled with the wormlike-micelle solution. Steel balls of diameters $d$ from 3 mm to 8 mm were introduced one at a time into the solution with zero initial velocity using an electromagnet positioned accurately to ensure that the balls sediment along the axis of the tube. Images of the falling balls were recorded using a Citius C100 Centurio camera capable of capturing 423 frames per second at its full resolution of 1280$\times$1024 pixels. Image analysis was done on the captured frames to extract information about the position and velocity of the ball as a function of time. Experimental runs were separated by at least 30 minutes in order to ensure complete relaxation and homogenization of the fluid prior to each run.

We present first our results for Gel A. Fig. \ref{bursts} shows velocity-time
plots of various diameters $d$. For $d=0.4 cm$ (Fig. \ref{bursts}(a)) the ball
quickly achieves a steady terminal velocity as expected in the Stokesian limit
for a Newtonian fluid. For $d=0.56 cm$ (Fig. \ref{bursts}(b)) a strong
oscillatory transient as in noted by King and Waters \cite{KingWaters} is seen
before terminal velocity sets in. For $d=0.70cm$ (Fig. \ref{bursts}(c)),
oscillations are seen to occur in repeated bursts on top of a constant baseline
velocity $v_{b}$, and are sustained throughout the fall of the particle. The
time between bursts is 4 to 7 sec, at least an order of magnitude larger than
the oscillation period $\sim 0.3$ s within the burst. Fig. \ref{bursts}(d) is an
enlarged version of the bursts highlighted in Fig. \ref{bursts}(c). Within each
burst the oscillations have an exponentially damped sinusoidal form; the total
number of such bursts before the ball reaches the bottom is too small for us to
conclude whether their timing is truly periodic or merely distributed narrowly
about a well-defined mean. The figures also show that the oscillations are
occasionally large enough to result in a small negative velocity; i.e., the ball
sometimes jumps up against gravity. Visual observation shows a strong negative
wake behind the ball during the burst. Gel B shows similar behaviour as seen in
Figs. \ref{bursts}(e-h). Interestingly, it should be noted that the baseline
velocities scale as $d^2$, consistent with Stokes's Law. Jayaraman \textit{et
al}.\cite{Belmonte} remarked that oscillations set in at velocity gradients
$v_{b}/d$ close to the frequency where $G''$ (loss modulus) begins to increase. We do not
observe such a correspondence.
\begin{figure}[h]
\centering
\includegraphics[height=8cm]{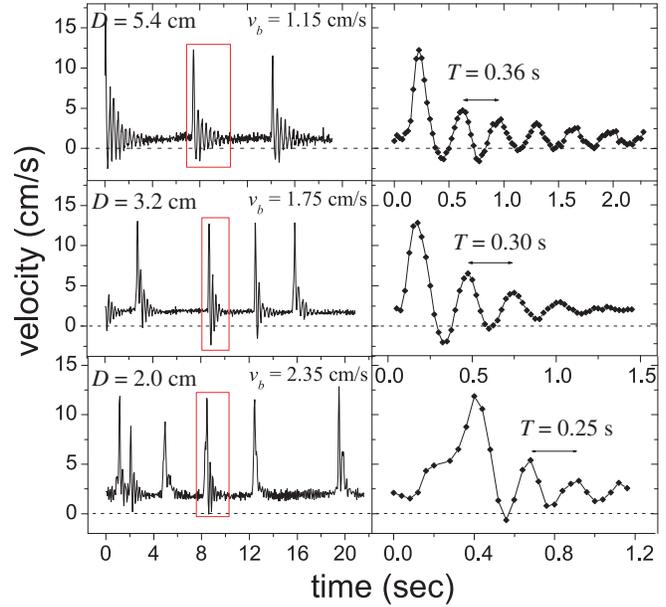}
\caption{The motion of 0.7 cm ball in CTAT (2wt$\%$)+100 mM NaCl dropped in tubes of various diameters. Note the increase in the baseline velocity, $v_{b}$ in thinner tubes (diameter $D$) followed by decreasing time-period ($T$) of burst oscillations.} \label{WallEffect}
\end{figure}

Fig. 2 shows the effect of wall confinement for a ball of diameter d=0.7 cm. A comparison of baseline velocities $v_{b}$ in tubes of diameter D = 5.4 cm, 3.2 cm and 2.0 cm shows that $v_{b}$ increases as D decreases, an effect opposite to that known in Newtonian fluids \cite{Chhabra}. The effect is observed for smaller diameters as well. We find that within a burst the time-period $T$ (as marked in Fig. \ref{WallEffect}) decreases with decreasing $D$. Local thinning caused by flow alignment of micellar structure along the walls, expected to be more pronounced in narrower tubes, is a likely explanation for these observations.

In order to account for the observation of bursts and oscillations, we now offer
a simple theoretical model based on the interplay of the orientation field of
the micelles and flow generated by the settling particle. We emphasize that our
intent is to offer a rational account of the observed phenomena, not to
reproduce them in full detail.
Further, we will show that the
model, although schematic in nature, nonetheless leads to other predictions that
can be tested in independent experiments. We start by summarizing some features
of the behaviour shown in Fig. \ref{bursts}(c). The phenomenon of primary
interest is the observation of rapidly oscillatory bursts repeated at fairly
regular long intervals. Each burst is damped in a roughly exponential manner,
with a time constant noticeably smaller than the interval between bursts, i.e.,
each successive burst emerges from a flow that has become nearly quiescent. The
baseline settling speed $v_0 \simeq 1.15$ cm/s, with oscillations up to 12 cm/s.
The resulting shear-rate at the scale of a particle (diameter $d \simeq 0.7$ cm)
is $\dot{\gamma} = v_0/d \simeq 1.64$ to $17.1$ s$^{-1}$, corresponding to a
timescale $\dot{\gamma}^{-1} \simeq .06$ s to $0.6$ s for the accumulation of a
strain of unity. The micellar solution has a measured zero-shear-rate viscosity
of $\eta_0 \simeq$ 18.0 Pa s, thinning under shear to about 1.4 Pa s, and a
viscoelastic or orientational relaxation time $\tau_v \simeq 1.34$ s. The
Reynolds number at the particle scale is thus at most $6 \times 10^{-4}$. The
settling particles are steel (density $\rho \simeq 8$ g/cm$^3$) and hence would
reach terminal velocity, if the viscosity were constant, in a time of order
$\tau_i = \rho d^2/18 \eta_0 \simeq$ 1.2 to 15.6 millisec depending on whether
one takes the zero-shear or thinned value for the viscosity. The observed period
\textit{within} the bursts is $T_1 \simeq 0.36$ s, and the period
\textit{between } bursts is $T_2 \simeq 9$ s. $T_1$ and $\tau_v$ are comparable
to $\dot{\gamma}^{-1}$, $T_2 \gg T_1, \tau_v$, and $\tau_i$ is much smaller than
all of these. Both acceleration and advection can thus be neglected in the
Navier-Stokes equation, and we can work in the limit of Stokesian hydrodynamics,
keeping track of stresses arising from the complex nature of the micellar
solution. We now present our model for the coupled dynamics of the particle and
the micellar medium. 

Consider a particle settling under gravity in a wormlike-micelle solution with velocity field $\mathbf{u(r)}$ at location $\mathbf{r}$, with solvent viscosity $\eta$ and a contribution $\mathbf{\sigma}^{(m)}$ to the stress tensor from the conformations of the micelles. The steady Stokes equation
\begin{equation}
\label{eq:stokes}
-\eta \nabla^2 \mathbf{u} = -\nabla P + \mathbf{W} -\nabla \cdot \sigma^{(m)}
\end{equation}
balances viscous, pressure, and conformational stresses against the
gravitational force from the buoyant weight $\mathbf{W}$ of the particle.
Inverting (\ref{eq:stokes}) and choosing the particle's centre of mass as
origin, the settling velocity
\begin{equation}
\label{eq:veleq1}
\mathbf{v} = \mathbf{v}_0 + \int d^3 r \, \mathbf{H(r)} \cdot \nabla \cdot \sigma^{(m)}(\mathbf{r})
\end{equation}
where $\mathbf{H}$ is the incompressible Stokesian hydrodynamic propagator and
$\mathbf{v}_0$ is the Stokes settling velocity in the absence of contributions
from micelle conformations, and the second term on the right in
(\ref{eq:veleq1}) contains modifications of the settling speed due to  micellar
stresses around the particle. The dynamics of $\sigma^{(m)}$ 
\begin{equation}
\label{eq:jslike}
{D \sigma^{(m)} \over Dt} = \lambda_0 \mathbf{A} + \lambda_1 [\mathbf{A} \cdot \sigma^{(m)}]_{ST} - {1 \over \tau} \sigma^{(m)}
\end{equation}
combines flow-orientation coupling and relaxation, familiar from a variety of contexts including general rheological models, rheochaos, and nematic hydrodynamics \cite{js,aradiancates,moumitaetalrheochaos,rienaecker}. Here $D/Dt$ is a time derivative comoving and corotating with the fluid, $\mathbf{A}$ is the symmetric deformation-rate tensor, and the subscript $ST$ denotes symmetric traceless part.

It is useful to work with a simple model extracted from (\ref{eq:veleq1}) and
(\ref{eq:jslike}). From (\ref{eq:veleq1}) we suggest that a component $p$ of the
micellar stress, presumably originating in micelle orientation around the
particle alters the settling speed: $v = v_0 + p$ where $v_0$ is the speed in
the absence of contributions from the micellar stress. The form of
(\ref{eq:jslike}) suggests the schematic dynamics $\dot{p} \sim v + v p + vq -
p+ ...$, where production by shear competes with relaxation to local
equilibrium. As in the Johnson-Segalman model \cite{js} or nematic hydrodynamics
\cite{degp,forster,rienaecker,moumitaetalrheochaos} shear, through $v$,
naturally couples $p$ to other components $q$, which in turn have a similar
dynamics. The ellipsis indicates terms of higher order in $p$ and $q$, and we
have assumed that the local shear-rate is proportional to the settling speed
$v$. The structure of these equations raises the possibility of oscillations
with a timescale set by the shear-rate. Such shear-induced mixing of micellar
stress components is very likely the mechanism for the oscillations
\textit{within} a burst, as their frequency, as seen from the numbers reported
above, lies right in the middle of the range of particle-scale shear-rates
$\dot{\gamma}$. The timescale associated with the repeated appearance of the
bursts is far longer, as remarked above, and must arise from a distinct process.
We propose that it enters through the memory mechanism introduced by
Cates and coworkers \cite{ajdarietal2002,aradiancates} that we incorporate and
explain in detail below. Assembling these ingredients we replace the dynamics of
the micellar stress tensor (\ref{eq:jslike}) by the effective model
\begin{equation}
\label{peq}
\dot{p} = \lambda_0 v + \lambda_1 v p + \lambda_2 v q  -\zeta p - s w -
\mathcal{P}(p,q),  
\end{equation}
\begin{equation}
\label{qeq}
\dot{q} = \mu_0 v + \mu_1 v p + \mu_2 v q   - \gamma q - \mathcal{Q}(p,q), 
\end{equation}
and (\ref{eq:veleq1}) by the schematic relation $v=v_0+p$ already mentioned
above. The terms $\mathcal{P}$ and $\mathcal{Q}$ in (\ref{peq}), (\ref{qeq})
incorporate nonlinearities that would arise naturally in a local thermodynamic
approach \cite{degp,moumitaetalrheochaos} where $p$ and $q$ are related to the
orientational order parameter. The coefficients $\lambda_i, \, \mu_i$ are
flow-orientation couplings, corresponding to similar terms in (\ref{eq:jslike})
and best understood from nematic hydrodynamics
\cite{forster,degp,rienaecker,starklubensky,moumitaetalrheochaos,lambdafootnote}
. We have allowed the relaxation of $p$ to occur
through two channels: locally in time via the term $\zeta p$ in
(\ref{peq}), and through the intervention of a structural variable\footnote{In
principle $q$ could also relax through such a memory
mechanism. For simplicity we have not explored this possibility.}
\cite{ajdarietal2002,aradiancates,fieldingOlmsted}
\begin{equation}
\label{memeq}
w(t) = \int_{-\infty}^t dt' e^{(t-t')\Gamma} p(t')
\end{equation}
determined by the past history of $p$. Equivalently, 
\begin{equation}
\label{weq}
\dot{w}=\Gamma(p-w),
\end{equation}
whence it is clear that $w$ becomes identical to $p$ only for $\Gamma \to
\infty$. We
see formally that $w$ and $p$ stand in the same relation as position and
momentum for an oscillator, even in regard to the relative signs of the
coefficients of $w$ in (\ref{peq}) and $p$ in (\ref{weq}). We emphasize that
this relation has nothing to do with real inertia: our equations are meant to
describe the viscosity-dominated regime. Equation (\ref{weq}) says, plausibly,
that an imposed nonzero local
alignment $p$ shifts the equilibrium value of an underlying structural quantity
such as the Maxwell viscoelastic time of the material
\cite{ajdarietal2002,aradiancates} or the micelle length \cite{fieldingOlmsted}.
As in those works, we assume that the parameters $\Gamma$ and $s$ of this
mechanochemical coupling are properties of the unsheared system; our
observations are consistent with $\Gamma \ll \dot{\gamma}$, the particle-scale
shear
rate.

\begin{figure}[t]
\centering
\includegraphics[height=6cm]{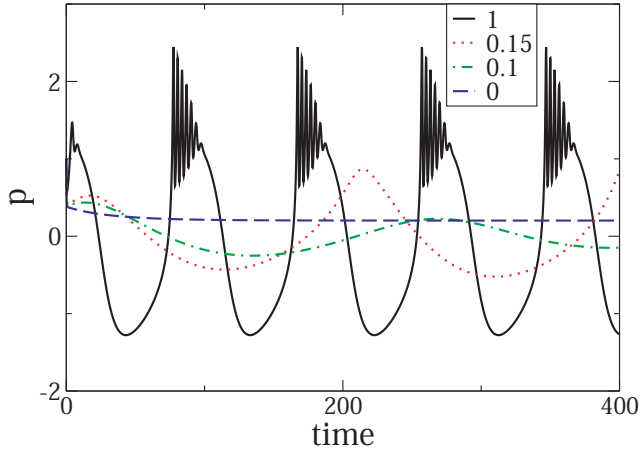}
\caption{Oscillations in the settling speed from (\ref{peq}), (\ref{qeq}), (\ref{weq}) for several $\alpha$ as indicated. Note the two distinct timescales, within and between bursts.}\label{OscReproduced}
\end{figure}

The parameter space of (\ref{peq}), (\ref{qeq}) and (\ref{weq}) is vast, and our limited exploration finds regimes of spontaneous oscillation. Fig. \ref{OscReproduced} shows $p$ (which is $v$ apart from a constant) as a function
of time for the model equations $\dot{p} = A_1 - A_2 p - A_3q - A_4 w + A_5p^2 - A_6p^3$, $\dot{q} = \alpha(B_1 - B_2 p - q + B_3 p^2)$, $\dot{w} =
\Gamma(p-w)$\footnote{with parameter values $A_1$=3, $A_2$=10, $A_3$=1,
$A_4$=$A_5$=6, $A_6$=1, $B_1$=3, $B_2$=12, $B_3$=6 \& $\Gamma$=0.006}, bringing
in
$q$ gradually by stepping $\alpha = 0,
0.1, 0.15 \& 1.0$.

Our model, constructed using physically plausible arguments, reproduces key
features of the experimental observation, namely, a rapid oscillation timescale
clearly associated with the shear and a much slower oscillation of distinct
origin.  Further, if the contribution of $q$ is diminished by reducing $\alpha$
only the slow oscillation of $p$ against $w$ is seen. Only when $\alpha$ is
large are the rapid oscillations seen, and these disappear on a timescale much
smaller than that associated with the slow oscillation.

As noted in \cite{aradiancates,bulbulprl2011}, equations of the general form
(\ref{peq}) - (\ref{weq}) are widespread in the neuronal modelling literature
\cite{fitz,hindmarsh}. Among the most robust features of such models are
oscillations with a bursty character, and the coexistence of two or more widely
separated timescales. In the neurophysics literature the introduction of an
additional slow ``recovery variable'' was motivated in part by detailed
knowledge of the electrophysiology of the axon \cite{hindmarsh}. The success of
our model in capturing essential features of our experiment encourages us to
suggest independent experimental tests for the coupled dynamics of micellar
stress $p$ and ``structure'' $w$ in the absence of shear. If not for the
structural variable $w$ we should have a conventional Johnson-Segalman or
nematogenic fluid \textit{at rest}, whose dynamical response and spontaneous
fluctuations are overdamped. However, in (\ref{peq}) and (\ref{weq}) the
micellar stress $p$ and the structural variable $w$ act like the momentum and
coordinate of an oscillator. And precisely because we require $w$ to be
\textit{slow}, the damping coefficient $\Gamma \ll s$, the coupled dynamics is
in the
\textit{underdamped} regime, so their linear response to disturbance must be
oscillatory. Further, if we augment the dynamical equations (\ref{peq}),
(\ref{weq}) with thermal noise, the resulting Fourier-transformed dynamic
correlation functions of the system in the absence of the falling ball should
show a peak at a frequency around that of the \textit{slow} oscillation seen in
the falling-ball experiments.This feature seems to be quite robust, relying only
on the assumption that the parameters in the dynamical equation for the
structural variable $w$ depend negligibly on the shear rate. This property
is shared by the models presented by Cates \textit{et al.}
\cite{ajdarietal2002,aradiancates} and emerges naturally from microscopic
kinetic models \cite{Cates_Fielding_review}. However, it is conceivable that
the kinetics in our system is such that $w$ becomes slow only under strong
imposed shear, in which case the \textit{equilibrium}, i.e., unsheared,
fluctuations of the coupled $p-w$ dynamics would be overdamped. A search for
such oscillatory response or correlation in \textit{equilibrium} experiments
would help resolve this issue. This in turn requires a dynamical probe of local
anisotropy; a possible scheme might be to suspend droplets of a strongly
flow-birefringent fluid in the wormlike-micelle solution, and measure the power
spectral density of the spontaneous fluctuations in the light intensity seen
through crossed polars.

We emphasize that our simple model aims primarily to capture the bursts
and the two disparate time-scales in the problem. Within the limitations of this
approach, we have suggested possible independent tests of the model. Further
experiments with different viscoelastic parameters and particle shapes
should lead us to a more quantitative theory, at the level
of reduced models like (\ref{peq}) - (\ref{weq}) or a more complete hydrodynamic
treatment. A better understanding, perhaps in terms of micelle kinetics
\cite{Cates_Fielding_review}, of the slow structural variable in our system is
of particular importance. 
 
To summarise: we have isolated the mechanism for the observed oscillatory
descent of a sphere through a wormlike-micelle solution. The oscillations
produced by (\ref{peq})-(\ref{weq}) show two broadly different timescales, as in
the experiments, although the detailed form of the oscillations is not identical
to those observed. One timescale is associated with the shear, the other with
the coupled dynamics of micellar stress and a structural parameter such as
micellar length or equilibrium Maxwell time. We argue that this second
timescale is likely to appear in the unsheared system in independent
measurements of the autocorrelation of spontaneous fluctuations of stress or
orientation, and suggest experiments to test this possibility.

Flow-birefringence studies \cite{rehage1986,tropea2007} focusing on the velocity
and
orientation fields of the medium around the falling ball will offer further
insight into the detailed mechanism underlying the phenomena presented in this
work. We note that the birefringence contrast as seen by Rehage \textit{et al.}
\cite{rehage1986}
in studies on solutions similar to ours, is small. Such measurements on our
system will therefore require a careful high resolution study, which we defer
to later work.  

Numerical studies of the complete hydrodynamic equations for
orientable
fluids, studying the effect of the gravitational settling of a sphere through
the medium, should see these curious oscillations, and will allow us to converge
on the correct effective model of the general type of (\ref{peq}) - (\ref{weq}).

We thank M.E. Cates for useful discussions. NK and SM are grateful to the
University Grants Commission, India for support. AKS and SR
respectively acknowledge a CSIR Bhatnagar fellowship and a DST J.C. Bose
fellowship.


\begin{thebibliography}{34}
\bibitem{lambdafootnote} In the underlying hydrodynamic equations there are only two independent couplings of this form, which enter both in the equation for $\sigma^{(m)}$ and in the generalized Stokes equation for the velocity. Equations (\ref{peq}) - (\ref{weq}) have been written in a form in which the coefficients in the Stokes equation have been absorbed into a rescaling of the orientation variables. We have allowed for distinct values for the flow-orientation couplings for $p$ and $q$ to take into account the possibility of detailed differences in orientation and flow fields at various locations around the particle.
\end{thebibliography}

\end{document}